\documentclass[runningheads]{llncs}
\usepackage[T1]{fontenc}
\usepackage{graphicx}
\usepackage{enumitem}
\usepackage{pdfpages}
\usepackage{subcaption}
\usepackage{tikz}
\usepackage{amsmath}
\usepackage{booktabs}
\usepackage{xcolor}
\usepackage{fancyhdr}
\usepackage[inkscapelatex=false]{svg}
\begin{document}
\title{Beyond Predictions: A Study of AI Strength and Weakness Transparency Communication on Human-AI Collaboration}
\titlerunning{A Study of AI Strength and Weakness Transparency Communication}
%
\author{Tina Behzad\inst{1}\thanks{Work done while interning at the Institute for Creative Technologies, University of Southern California.}\orcidID{0009-0009-1157-9082} \and
Nikolos Gurney\inst{2}\orcidID{0000-0003-3479-2037} \and
Ning Wang\inst{2}\orcidID{0009-0004-7037-9260} \and David V. Pynadath\inst{3}\orcidID{0000-0003-2452-4733}}
\authorrunning{T. Behzad et al.}
%
\institute{Stony Brook University, Stony Brook, NY
\email{tbehzad@cs.stonybrook.edu}\\\and
Insitute for Creative Technologies, University of Southern California, Los Angeles, CA
\email{\{gurney,	nwang\}@ict.usc.edu}\\ \and
Rice University, Houston, TX\\
\email{pynadath@rice.edu}}

\fancyfoot[C]{\small Preprint. Final version to appear in LNCS, Springer.}
\pagestyle{fancy}
\maketitle   
%

\renewcommand\thefootnote{}\footnotetext{
The final authenticated version will be published in Lecture Notes in Computer Science (LNCS), Springer.
}
\addtocounter{footnote}{-1}
\begin{abstract}
The promise of human-AI teaming lies in humans and AI working together to achieve performance levels neither could accomplish alone. Effective communication between AI and humans is crucial for teamwork, enabling users to efficiently benefit from AI assistance. This paper investigates how AI communication impacts human-AI team performance. We examine AI explanations that convey an awareness of its strengths and limitations. To achieve this, we train a decision tree on the model’s mistakes, allowing it to recognize and explain where and why it might err. Through a user study on an income prediction task, we assess the impact of varying levels of information and explanations about AI predictions. Our results show that AI performance insights enhance task performance, and conveying AI awareness of its strengths and weaknesses improves trust calibration. These findings highlight the importance of considering how information delivery influences user trust and reliance in AI-assisted decision-making.

\end{abstract}

\section{Introduction}
With recent advancements in the quality and accessibility of Artificial Intelligence (AI), these systems are becoming increasingly integrated into society. In response, policymakers and practitioners emphasize the need for greater human oversight in AI-driven decision-making \cite{passi2022overreliance}. This shift necessitates human-AI collaboration, where individuals must review AI recommendations and ultimately make the final decision. In such scenarios, human's trust in their AI decision-aid becomes critical for the team's success \cite{glikson2020human}. The human decision-makers need to know when to trust or distrust an AI model's recommendations. Decades of research on this topic yielded complex insights into humans’ inclination to trust algorithms \cite{zhang2020effect} and the problematic disuse or overuse of automation \cite{lee2008review}. Numerous studies have demonstrated that individuals frequently avoid relying on decision-making systems in various scenarios \cite{bigman2018people,yeomans2019making}. However, it has also been shown that in many situations, people do prefer algorithms, making overreliance a sensible worry \cite{logg2019algorithm}.

Research on trust calibration in AI can be broadly categorized into two main approaches. The first approach emphasizes explainability, suggesting that making black-box models more interpretable will help users calibrate their trust in AI systems \cite{yeomans2019making,ribeiro2016should}. However, several recent empirical studies have found little evidence that higher explainability significantly impacts users' willingness to trust machine learning models \cite{cheng2019explaining,kunkel2019let}. The second approach advocates for providing high-level information about the AI system, such as its accuracy, to help users adjust their trust levels accordingly \cite{lu2021human}. However, presenting high-performance metrics uncritically can sometimes lead to overreliance, where users place excessive trust in AI recommendations \cite{lai2019human}.

These findings highlight the level of information provided to the user has significant impact on user's trust \cite{passi2022overreliance}.  Overwhelming users with too much information about the model leads to user following both correct and incorrect decisions more often \cite{suresh2020misplaced} while the correct level of information is proven to be helpful \cite{de2020case}. The AI needs to communicate its own strength and weakness. By acknowledging its limitations, the AI can provide high-level explanations for its decisions, making it easier for humans to understand and verify its reasoning.

In this study, we examine how varying levels of information about an AI system’s performance, reflecting different degrees of awareness of its own limitations, affect human-AI collaboration in terms of performance, trust, and understanding. To investigate this, we develop a self-assessing AI model, drawing inspiration from explainability methods such as Local Interpretable Model-agnostic Explanations (LIME) \cite{ribeiro2016should}. We then conduct an empirical evaluation with 272 participants recruited through Prolific, testing our hypothesis on whether and how different levels of information and feedback impact users’ trust and decision-making performance over an income prediction task. Our results show that providing any level of information on AI's performance improves overall task performance compared to having no feedback. This finding aligns with prior research which suggests performance insights can enhance human-AI collaboration \cite{lai2020chicago}. Regarding trust calibration, our findings indicate that providing AI's confidence in its decisions or awareness of its strengths and weaknesses, both enable users to better discern when to trust the model's predictions. Moreover, conveying AI's awareness compared to confidence appears to slightly enhance users' ability to calibrate their trust more effectively.  

These findings highlight the importance of carefully designing the way AI communicates performance information to users. Given the diverse and sometimes contradictory research on what factors influence trust in AI, it is crucial to investigate the granularity of information provided and how it is conveyed. Our work underscores the need for further research on the impact of different levels and formats of feedback, helping to refine human-AI interaction strategies and ensure users can develop appropriately calibrated trust in AI systems.

\section{Related Work}
Advances in Machine Learning (ML) and ML-based AI in recent years have enabled these systems to exceed human-level performance in making predictions. One particularly important use case of machine learning is supporting decisions. Decision support systems (DSS) have undergone several evolutionary waves, and the integration of machine learning promises to drive another significant leap forward \cite{watson2017preparing}. However, despite this progress, algorithmic aversion, where people distrust and avoid using algorithms for decision-making, especially after observing them make mistakes \cite{dietvorst2015algorithm}, has become a major barrier to fully leveraging their capabilities. Research shows that users often resist incorporating algorithms into decision-making across various domains, including critical tasks such as aiding professionals in making medical recommendations \cite{promberger2006patients,shaffer2013patients}, receiving medical \cite{logg2019algorithm} or financial \cite{eastwood2012people} advice, and employee selection \cite{diab2011lay}. This resistance even extends to lower-stakes tasks, such as receiving joke recommendations \cite{yeomans2019making}, raising concerns about the feasibility of joint human-algorithm decision-making in practice \cite{burton2020systematic}.

Previous research has tried to understand and address the factors contributing to this reluctance and distrust.  A review of studies on the topic between 1950 and 2018 identified key themes influencing algorithmic aversion, including expectations and expertise, decision autonomy, incentivization, cognitive compatibility, and divergent rationalities \cite{burton2020systematic}. 

However, while some users resist AI-based decisions, others display overreliance on AI, accepting its recommendations even when they are incorrect. This overreliance could be caused by human decision-making biases, such as automation bias \cite{pop2015individual,logg2019algorithm} and confirmation bias \cite{lu2021human}. Such overreliance is also influenced by various human factors, e.g., individual differences \cite{passi2022overreliance,pop2015individual,chong2022human}, and situational factors, like ordering effects, such as the sequence in which AI errors occur\cite{nourani2020role,nourani2021anchoring}.

This often occurs when users struggle to assess whether—and to what extent—they should trust the AI \cite{passi2022overreliance}, highlighting that achieving high trust as a solution to algorithmic aversion should not be the ultimate goal. Insufficient trust may lead users to reject AI assistance, even when it could improve outcomes (distrust/aversion) \cite{leichtmann2023effects}. Conversely, excessive trust can cause users to perform worse than either the AI or human alone \cite{bansal2021does}. An appropriate level of trust—calibrated trust—  is essential \cite{kraus2020more}. As a result, trust calibration has emerged as a critical area of research.

Trust calibration has been extensively studied in automation tasks \cite{10.1145/2516540.2516554,lee2004trust,mcguirl2006supporting} and more recently in AI-assisted decision-making \cite{lai2019human,schaffer2019can,yin2019understanding}. Several studies have examined the impact of accuracy information on trust, showing that users tend to increase their trust in AI when high accuracy indicators are displayed \cite{yin2019understanding,rechkemmer2022confidence}. Others have focused on the role of explanations in shaping trust, arguing that the black-box nature of AI presents a barrier to adoption \cite{bhatt2019building,leichtmann2023effects,10.1145/3397481.3450650}. More recently, Daehwan Ahn et al. \cite{10.1145/3613904.3642780} integrated these two research streams to examine their relationship. Their findings indicate that while both factors had modest effects on participants' performance, interpretability did not lead to a robust improvement in trust, whereas providing accuracy information significantly increased trust. Building on previous research highlighting the significance and impact of outcome feedback, in this paper, we investigate how varying levels of feedback influence human-AI collaboration.
\section{Self-assessing AI} \label{sec:flaw-aware}
Exploring different levels of outcome feedback, our goal was to develop a model that is aware of its strengths and limitations and capable of identifying predictions where it might be incorrect. A straightforward and widely used approach to achieve this is through confidence scores, which provide a measure of uncertainty.

The method for calculating these scores varies depending on the model type. For probabilistic models such as Naive Bayes and Logistic Regression, confidence scores are directly obtained as class probabilities produced by the model. In ensemble models like Random Forests or Gradient Boosting, confidence scores can be calculated as the proportion of trees/models voting for a specific class. For more complex models, such as Neural Networks, the confidence score is typically derived from the activation function of the output layer, such as the softmax function, which provides a probability distribution over classes\cite{vemuri2020scoring}.

Confidence scores have been explored as a way to provide outcome feedback to users. Zhang et al. found that confidence scores help calibrate users' trust in AI models \cite{zhang2020effect}. Similarly, Rechkemmer et al. demonstrated that a model's confidence level significantly influences users' perception of its accuracy, impacting both their willingness to follow its predictions and their self-reported trust in the model \cite{rechkemmer2022confidence}. 

While confidence scores shows promise in guiding human-AI interaction, they do not constitute awareness—they are simply mathematical outputs derived from the model’s internal calculations. These scores do not imply that the model \textbf{knows why} it might be wrong; rather, they reflect the model’s certainty based on its training data without any underlying reasoning ability. To address these limitations, alternative approaches have been explored. One promising
approach is evidential learning \cite{sensoy2018evidential,sensoy2020uncertainty}. The evidential learning approach learns a generative model to create out-of distribution samples so that the classifier can be explicitly taught
the input regions it should be uncertain about \cite{sensoy2020uncertainty}. However, this approach introduces additional complexities in the training process \cite{tomsett2020rapid}.

Another important consideration is how humans process probabilistic information. Research has shown that people often struggle with probabilistic reasoning, falling into errors such as the base-rate fallacy \cite{kahneman1982variants,tversky1982judgment,tversky1981framing}. Studies suggest that alternative, non-probabilistic representations of uncertainty or confidence can lead to improved trust calibration \cite{helldin2013presenting,mcguirl2006supporting}.
Building on these findings and inspired by similar approaches in explainable AI, such as Local Interpretable Model-agnostic Explanations (LIME) \cite{ribeiro2016should}, which train simpler, more interpretable models to approximate the behavior of complex models, we adopt a similar strategy. Specifically, we train a decision tree, a highly interpretable model, on the original dataset but with labels indicating whether the predictions of the original model were correct or incorrect. This allows us to create a model that not only identifies where the complex model is likely to make mistakes but also explains why, by analyzing how different features contribute to these errors. Additionally, the decision tree can predict for new data points whether the original model is likely to make a mistake.

\subsection{Defining the Task} \label{sec:flaw-aware-task}

Researchers have argued that results and recommendations for human-centered interaction with AI may vary depending on the context of use \cite{leichtmann2023effects}. For our study, we selected an income prediction task, which has been used in prior user studies on decision support systems \cite{vodrahalli2022humans}. This scenario was chosen based on the criteria outlined by Leichtmann et al. \cite{leichtmann2023effects}. Income prediction is highly \textbf{relevant}, as it plays a crucial role in hiring, financial assessments, and social policy. Unlike specialized fields like medical diagnosis, income-related decisions \textbf{do not require expert knowledge}, making it easier to recruit participants from diverse backgrounds. Additionally, people regularly assess income-related factors in everyday life, ensuring the task is \textbf{close to participants' reality}. Finally, concerns surrounding income equity and fairness make this a topic of significant \textbf{public interest}.

We used the American Community Survey (ACS) Public Use Microdata Sample (PUMS) dataset \cite{ding2021retiring}, which covers multiple years and all states across the United States. It supports five different prediction tasks, including income prediction. For this study, we restricted the dataset to individuals from California in the year 2018, resulting in $196,665$ individuals.

We chose our original model to be a Random Forest classifier, trained using the scikit-learn library \cite{scikit-learn} with default hyperparameters. We used 70\% of the data for training, achieving an accuracy of 80\% on the test set. To create a self-assessing system, we trained a decision tree on the same training dataset but replaced the original labels with binary indicators of whether the original model’s predictions were correct. We selected a decision tree model for three key reasons. First, its structure is inherently interpretable, allowing us to trace and explain each decision path. Second, because decision trees align with everyday reasoning, a lot of people without any background can understand the model’s logic, and we can tailor the depth of information displayed to the user. Third, the tree’s natural grouping of data points into branches creates clusters that we leverage later in our study design (see section \ref{sec:study-design}). Since the original model had an accuracy of 90\% on the training data, the resulting dataset was highly imbalanced. To address this, we balanced the data to contain an equal number of correct and incorrect labels before training the decision tree. After tuning hyperparameters, the self-assessing model achieved a 66\% accuracy which was the best we could get under these conditions. We call this tree the flaw decision tree\footnote{To have a better understanding of how the tree looks, refer to Figure \ref{fig:tree}.}.

\section{Methods}
\subsection{Study Design} \label{sec:study-design}
Our main hypotheses when designing the experiments were:
\begin{enumerate}[label=\textbf{H\arabic*}, ref=H\arabic*]
    \item \textit{Communicating awareness of the model's weakness and strength can improve task performance.} \label{hyp:H1} 
    \item \textit{Communicating awareness of model's weakness and strength can help human teammates calibrate trust in AI.} \label{hyp:H2}
    \item \textit{Communicating awareness of the model's weakness and strength can help human teammates understand AI better.} \label{hyp:H3}
\end{enumerate}

To evaluate this, we designed an experiment consisting of four treatment conditions and a control group. The task required participants to predict whether an individual's income would be above or below \$80K after reviewing relevant attributes. All groups were presented with the individual's information in a tabular format, as shown in Figure \ref{fig:groups}. The treatment groups received varying levels of assistance and feedback from an AI teammate (the trained random forest predictor from section \ref{sec:flaw-aware}), while the control group (we also call this group \textbf{No Helper} through the text) completed the task without AI support. This setup allowed us to assess participants' performance both without and with different levels of AI assistance and outcome feedback.

The four treatment groups were as follows (illustrated in Figure \ref{fig:groups} in appendix \ref{appendix:study-design}):
\begin{enumerate}[label=\textbf{G\arabic*}]
     \item \textbf{Decision-Only}: Saw the AI helper’s prediction for each individual alongside the individual's information. 
     \item \textbf{Decision + 80\% Exp}: Received the same information as \textbf{G1}, but before starting, they were informed that the AI’s accuracy is 80\% and were given an explanation of what this means.
     \item \textbf{Decision + Confidence Exp}: In addition to seeing the AI's prediction and the individual's attributes, they were also shown the model’s confidence in its prediction.
     \item \textbf{Decision + self-assessing Exp}: Along with the AI's prediction and the individual's attributes, they were provided with information on how well the model performs for similar data points and the overall performance (80\%). 
\end{enumerate}

We selected 50 data points from the full test dataset described in section \ref{sec:flaw-aware-task}, allocating 10 for the training phase and 40 for the main task. To ensure consistency, we selected instances in a way that the AI's predictions maintained the reported 80\% accuracy (2 of 10 incorrect predictions in training and 8 out of 40 predictions in the main task). 

For Group 3, the confidence score corresponded to the probability of the predicted class (income level above \$80K). As stated in the scikit-learn documentation, this score is calculated as the mean predicted class probabilities of the trees in the forest \footnote{The class probability of a single tree is determined by the fraction of samples of the same class in a leaf.}. For Group 4, we use the self-assessing model described in section \ref{sec:flaw-aware}. Initially, we considered presenting participants with the entire flaw decision tree, highlighting where each individual’s data point falls within the tree or displaying the full root-to-leaf decision path. However, we determined that this approach could overwhelm participants with too much information, particularly for those unfamiliar with tree structures. We chose to use the flaw decision tree to categorize data points into groups, where each leaf node represents a group of similar individuals based on whether the original model predicts them correctly. For each group, we report the fraction of individuals that were predicted correctly by the original model. We present this information as "AI’s accuracy for similar individuals", alongside the general statement that the AI’s overall accuracy is 80\% for each prediction.

When selecting the 50 data points from the test dataset, we ensured that the additional information provided to treatment groups 3 and 4 remained consistent. Specifically, if the confidence score displayed to Group 3 for a given point was high, the AI’s accuracy for similar individuals shown to Group 4 was also high, and vice versa. While the exact numerical values may differ, we ensured that both metrics consistently reflected whether they were above or below the overall 80\% accuracy threshold. Additionally, for instances where the original model made incorrect predictions, we balanced the selection of data points. Half of these instances were chosen so that the reported confidence or accuracy was informative (i.e., lower than the overall 80\% accuracy), indicating appropriate uncertainty. The other half was selected to be misleading, where the model was overconfident despite being incorrect.

For each of the aforementioned conditions, after identifying the subset of data corresponding to that specific condition, the points were randomly selected from the entire test set.


\subsection{Ethical Approval}
This study was approved by the University of Southern California Institutional Review Board.

\subsection{Recruitment}
We recruited 272 (approximately 50 per condition) US-based participants using the Prolific pool of study participants, where no personally identifying information was accessible. They were informed that the study would take place online, might involve working with an AI, last approximately fifteen minutes, pay \$4 plus a bonus of up to \$4, and was open to adult US citizens (a funding constraint). The average participant was 37 years old, most of whom ($n = 177$) reported being female. The modal reported level of education was a bachelor's degree ($n = 102$); the next most common was a high school diploma or equivalent ($n = 74$). The majority of participants ($n = 185$) self identified as white. The median completion time was 889.5 seconds, however, the completion time data are characterized by a right skew ($mean = 1025.2$ seconds). 

\subsection{Procedure}
We collected data in two batches: the four treatment conditions and the control group, which did not have an AI helper. After accepting the task on Prolific, the participant was redirected to a survey-based platform (Qualtrics) where they signed the consent form and agreed to participate in the study.
The survey software randomized treatment condition participants into a study arm. Verification of their Prolific ID advanced participants to an introduction page that introduced them to the task and, if in a treatment condition, how and what the AI helper would communicate to them. Next, they progressed to a training phase in which they were allowed to do 10 practice tasks that did not impact their bonus payment. Participants in treatment conditions received the same help from the AI that they eventually did during the incentivized portion of the study. The training phase ended with a report telling participants how well they did in the classification task and reminding them that the following 40 classifications tasks were bonus-eligible. In both the training and the actual task, users received feedback on whether or not they made the correct decision after each decision. At the end of the 40 tasks, participants were informed of their performance and the bonus amount they earned. 

Participants next completed a set of self-report questions (see appendix \ref{appendix:survey} for the complete set of questions). The Generalized Attitudes Towards AI Scale \cite{SCHEPMAN2020100014} followed the self-report. Finally, participants completed the demographics portion of the study, after which they were redirected back to Prolific to complete the platform requirements for payment. 

\subsection{Measures} \label{sec:measures}
\paragraph{Task Performance}: We measure performance as the proportion of correct final decisions, computed as the number of accurate predictions divided by the total number of predictions (40).
\paragraph{Compliance}: Compliance is defined as instances where the user's final decision is aligned with the AI's recommendation, serving as a proxy for trust in the AI’s decision. To further investigate users' ability to calibrate their trust, we differentiate compliance rates based on the correctness and confidence of the AI's predictions. Specifically, we distinguish between cases where the AI was (i) correct, (ii) incorrect and overconfident, and (iii) incorrect with an appropriate level of confidence. This differentiation allows us to assess whether users could appropriately override AI recommendations when the model was incorrect.
\paragraph{Self-Reported Perceptions of the AI Assistant}: To evaluate whether varying levels of outcome feedback influenced participants' understanding of the AI teammate, we administered a post-task questionnaire consisting of four key questions. Participants rated their agreement with each statement using a slider scale from 0 to 100. These questions were not presented to the control group. 1) The AI understands how the information in the tables relates to income levels. 2)The AI knows its own limitations. 3) I trusted the AI to provide useful suggestions. 4) I am confident that I know how the AI makes its suggestions.




\section{Results}
In the following, we present results related to our first (\ref{hyp:H1}) and second (\ref{hyp:H2}) hypotheses. Our self-reported measures on participants' understanding of the AI helper (\ref{hyp:H3}) showed no significant differences across groups. Due to space constraints, a more detailed discussion of these findings has been moved to the appendix.

\subsection{Task Performance}
\begin{figure}
    \centering
    \includegraphics[width=0.8\linewidth]{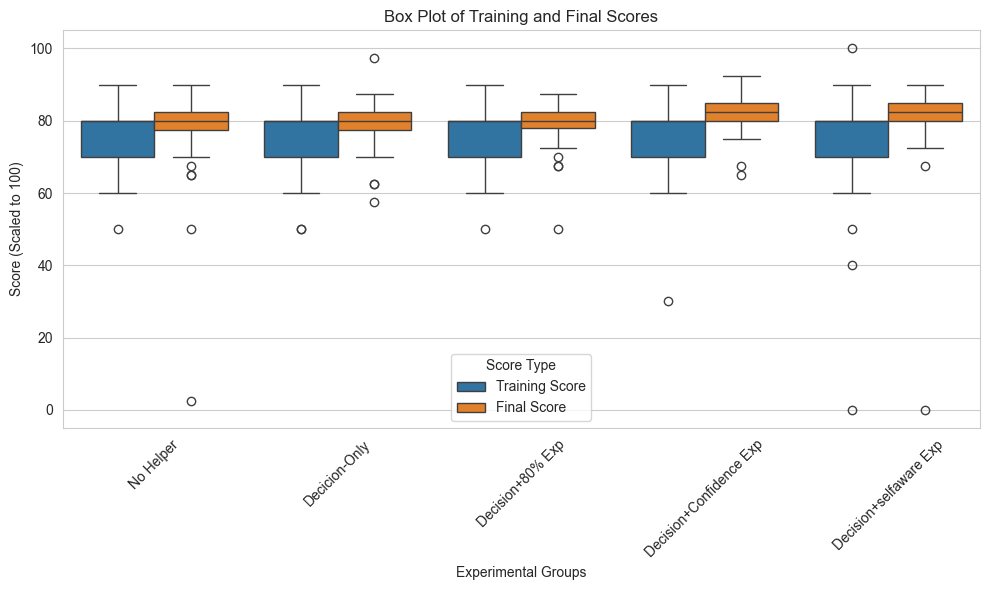}
    \caption{Box plot illustrating training scores (blue) and task scores (orange) for each group.}
    \label{fig:group-scores}
\end{figure}

We start by looking at task performance. Figure \ref{fig:group-scores} shows participants' scores across the training phase and for the actual task. The scores were scaled from to 100 for illustration (exact mean value can be found in table \ref{table:scores} in appendix \ref{appendix:results}). 
To better interpret the results we fitted a linear regression model(presented in Appendix \ref{appendix:results}, Table \ref{table: linear mod of score}). Results suggest that participants with input from an AI helper generally did better than participants in the \textit{No Helper}, i.e., control, condition (see Table \ref{table: linear mod of score} column (1) for the values used in the following interpretation). The average score of the \textit{No Helper} participants was 0.780, or just over 31 correct classifications. Participants in the \textit{Decision-Only} and \textit{Decision + 80\% Exp} conditions did better (their average scores were $0.780 + 0.010 = 0.790$ and $0.780 + 0.016 = 0.796$, respectively), although not significantly. Participants in the other two conditions \textit{Decision + Confidence Exp} and \textit{Decision + self-assessing Exp} did significantly better than the control participants, averaging $0.780 + 0.045 = 0.825$ and $0.780 + 0.035 = 0.815$, respectively---equating to roughly 33 correct classifications. 

The effects remain roughly the same when adding control variables to the basic model (see Table \ref{table: linear mod of score} column (2) for the values used in the following interpretation). We reached this model through iteratively specifying models using the various demographic and control variables (e.g, \textit{Training Score}, \textit{Age}, \textit{Sex}, \textit{Education}, etc.) and using a $\chi^2$ test to see if the reduction in the residual sum of squares was justified by the higher complexity of a model with additional variables. Generally speaking, participants that did better during the training phase did better during the incentivized task, and each additional year of age was associated with a small but significant decrease in score.

\subsection{Compliance}
Although predicting participants' overall performance (their score) in the task is valuable, it is arguably more important to understand how calibrated they are when they should heed or ignore the helper's advice. Well-calibrated compliance is not only correlated with performance but also opinions of the help \cite{gurney2023my}. For this, we use logistic regression models and, having established the importance of \textit{Age} and \textit{Training Score} in our previous efforts, only discuss the fully specified models (see Table \ref{table: logistic compliance models}). Note that we do not include the data from the participants in the \textit{No Helper} treatment condition as they were not choosing whether to comply with the recommendation of an AI helper. Thus, the \textit{Decision-Only} treatment condition serves as the ``control'' condition. 

Our results show that relative to the \textit{Decision-Only}, participants in each of the other three helper conditions were significantly more likely to comply. When we look at overall compliance, meaning over the entire task regardless of whether or not the helper was correct, the strongest impact is observed in \textit{Decision + Confidence Exp} condition which was associated with a 0.305 high log odds of complying (column (1), Table \ref{table: logistic compliance models}). In other words, for these participants, there was a 35\% increase in the odds of complying ($(e^{0.305}-1)*100)$. However, when we look at compliance when the helper was correct (column (3), Table \ref{table: logistic compliance models}), we see the most substantial effect appears in the \textit{Decision + self-assessing Exp} group increase in the odds of complying, followed by \textit{Decision + Confidence Exp}.

We further examine the odds of compliance when the AI is incorrect and overconfident versus incorrect with the right level of confidence (see Table \ref{tab:compliance incorrect}). Participants in the \textit{Decision + Confidence Exp} and \textit{Decision + self-assessing Exp} conditions are significantly more likely to comply when the AI is incorrect but overconfident and less likely to comply when the AI is incorrect but calibrated in its confidence. While the differences between the two groups are not statistically significant in either scenario, we observe slightly lower compliance in the \textit{Decision + self-assessing Exp} rate when the AI is overconfident (and similarly for the right confidence case), suggesting that participants in the awareness condition demonstrated better trust calibration in AI’s decisions.
\section{Discussion}
In this paper, we explored the design of an AI decision aid that generates explanations highlighting its strengths and weaknesses in predicting income levels. Our hypothesis posited that providing concise, awareness-based insights—reflecting a deeper understanding of the model’s limitations—could enhance task performance, trust calibration, and users’ comprehension of their AI teammate. To evaluate this, we conducted a user study with 272 participants, including a control group and four treatment conditions, each offering varying levels of assistance and feedback from the AI teammate. Our results demonstrate that providing AI performance information enhances task performance and that conveying AI’s awareness of its strengths and weaknesses helps users calibrate their trust in its decisions slightly better compared to showing only confidence levels. 

Our findings align with prior research suggesting that providing information about AI models helps users develop appropriate reliance on AI while also offering insights into how to enhance the effectiveness of this information. Overreliance is particularly critical in cases where the AI is incorrect, yet users choose to follow its recommendations. Our results suggest that conveying AI awareness can slightly reduce this blind trust more effectively than simply providing confidence levels.

While we did not observe significant differences across groups in terms of their understanding of AI, this may be due to limitations in the objective measures used. Future research could address this by incorporating follow-up assessments to more effectively evaluate participants' understanding of AI. Additionally, it would be valuable to examine whether these findings replicate across different tasks and varying levels of AI accuracy. Insights from such studies can further illuminate the complex dynamics of human-AI collaboration.
\section{Acknowledgement}
Research was sponsored by the Army Research Office and was accomplished under Cooperative Agreement Number W911NF-20-2-0053. The views and conclusions contained in this document are those of the authors and should not be interpreted as representing the official policies, either expressed or implied, of the Army Research Office or the U.S. Government. The U.S. Government is authorized to reproduce and distribute reprints for Government purposes notwithstanding any copyright notation herein.
\bibliographystyle{splncs04}
\bibliography{mybibliography}

\appendix
\section{self-assessing AI}
\begin{figure}[h]
    \centering
    \includegraphics[width=0.8\linewidth]{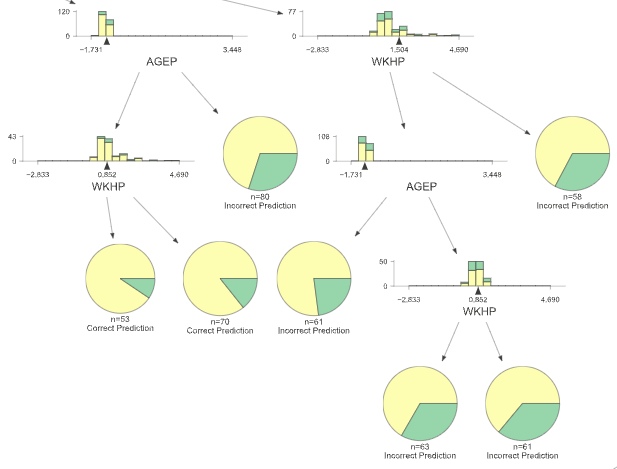}
    \caption{A small selection of leaf nodes from the self-assessing decision tree trained (values for features are normalized).}
    \label{fig:tree}
\end{figure}

A subset of the decision tree trained on the initial model, as described in section \ref{sec:flaw-aware-task}, is presented in Figure \ref{fig:tree}. The full tree has a depth of 10 and consists of 517 nodes.

\section{Study Design} \label{appendix:study-design}
\begin{figure}[h]
    \centering
    \begin{minipage}{0.48\textwidth}  
        \centering
        \includegraphics[height=12cm]{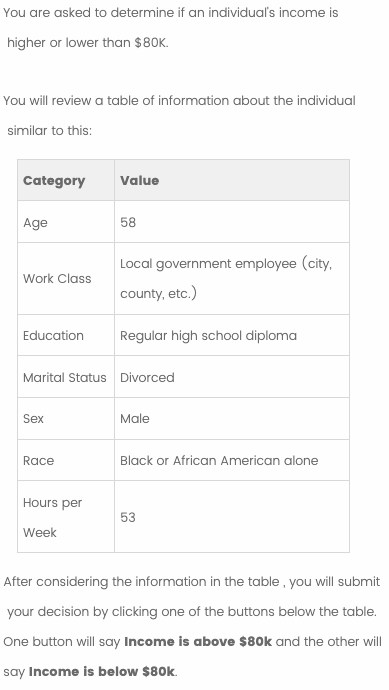}  
        \caption{The information provided to all groups.}
        \label{fig:left}
    \end{minipage}
    \hfill
    \begin{minipage}{0.48\textwidth}  
        \centering
        \begin{subfigure}{\textwidth}
            \centering
            \includegraphics[width=\linewidth]{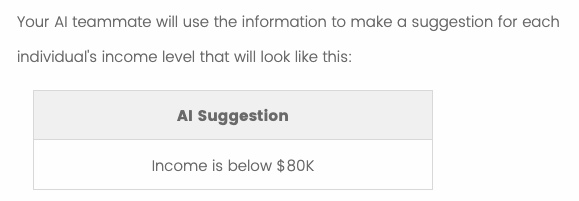}
            \caption{Additional information shown to \textbf{G2}}
            \label{fig:right1}
        \end{subfigure}
        
        \begin{subfigure}{\textwidth}
            \centering
            \includegraphics[width=\linewidth]{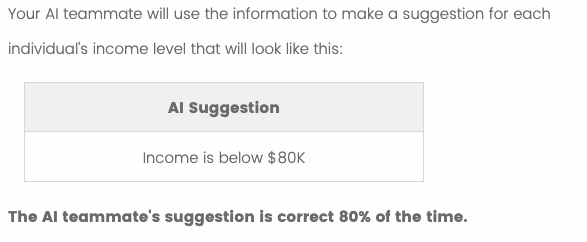}
            \caption{Additional information shown to \textbf{G3}}
            \label{fig:right2}
        \end{subfigure}
        
        \begin{subfigure}{\textwidth}
            \centering
            \includegraphics[width=\linewidth]{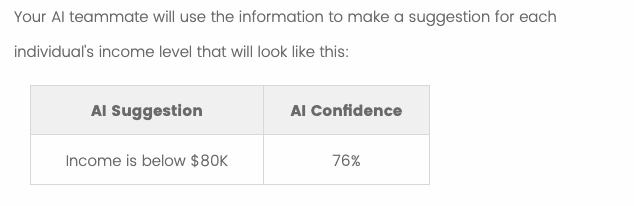}
            \caption{Additional information shown to \textbf{G4}}
            \label{fig:right3}
        \end{subfigure}
        
        \begin{subfigure}{\textwidth}
            \centering
            \includegraphics[width=\linewidth]{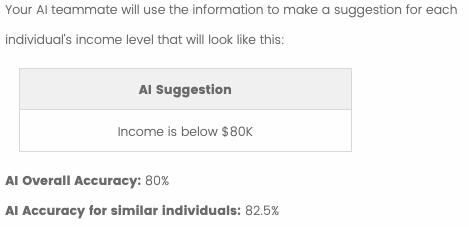}
            \caption{Additional information shown to \textbf{G5}}
            \label{fig:right4}
        \end{subfigure}
    \end{minipage}
    
    \caption{Example of the introduction shown to all participants}
    \label{fig:groups}
\end{figure}

Figure \ref{fig:groups} shows different conditions described in section \ref{sec:study-design}. Figure \ref{fig:left} shows the information provided to all participants and Figures \ref{fig:right1} to \ref{fig:right4} show different levels of information provided to different groups.

\section{Results}\label{appendix:results}

\begin{table}[h] \centering 
\caption{Mean Scores by Condition}
  \label{table:scores} 
\begin{tabular}[t]{lr}
\toprule
Condition & Mean Score\\
\midrule
No Helper & 0.780\\
Decision-Only & 0.790\\
Decision + 80\% Exp & 0.796\\
Decision + Confidence Exp & 0.825\\
Decision + self-assessing Exp & 0.815\\
\bottomrule
\end{tabular}
\end{table}
\begin{table}[h] \centering 
  \caption{Column (1) presents a linear regression model predicting a Score (number of correct income classifications divided by total classifications). The intercept, or constant, is the \textit{No Helper} condition. Column (2) adds control variables for the participants' performance during the training stage and age.} 
  \label{table: linear mod of score} 
\begin{tabular}{@{\extracolsep{5pt}}lcc} 
\\[-1.8ex]\hline 
\hline \\[-1.8ex] 
 & \multicolumn{2}{c}{\textit{Dependent variable: Score}} \\ 
\cline{2-3} 
\\[-1.8ex] & \multicolumn{2}{c}{$n$ Correct $\div$ Total} \\ 
\\[-1.8ex] & (1) & (2)\\ 
\hline \\[-1.8ex] 
 Decision-Only & 0.010 & 0.011 \\ 
  & (0.017) & (0.016) \\ 
  & & \\ 
 Decision + 80\% Exp & 0.016 & 0.011 \\ 
  & (0.017) & (0.016) \\ 
  & & \\ 
 Decision + Confidence Exp & 0.045$^{**}$ & 0.041$^{*}$ \\ 
  & (0.017) & (0.016) \\ 
  & & \\ 
 Decision + self-assessing Exp & 0.035$^{*}$ & 0.037$^{*}$ \\ 
  & (0.017) & (0.016) \\ 
  & & \\ 
 Training Score &  & 0.038$^{***}$ \\ 
  &  & (0.005) \\ 
  & & \\ 
 Age &  & $-$0.001$^{*}$ \\ 
  &  & (0.0004) \\ 
  & & \\ 
 Constant (No Helper) & 0.780$^{***}$ & 0.526$^{***}$ \\ 
  & (0.012) & (0.041) \\ 
  & & \\ 
\hline \\[-1.8ex] 
Observations & 272 & 272 \\ 
R$^{2}$ & 0.033 & 0.222 \\ 
Adjusted R$^{2}$ & 0.018 & 0.204 \\ 
Residual Std. Error & 0.090 (df = 267) & 0.081 (df = 265) \\ 
F Statistic & 2.257 (df = 4; 267) & 12.575$^{***}$ (df = 6; 265) \\ 
\hline 
\hline \\[-1.8ex] 
\textit{Note:}  & \multicolumn{2}{r}{$^{*}$p$<$0.05; $^{**}$p$<$0.01; $^{***}$p$<$0.001} \\ 
\end{tabular} 
\end{table} 

\begin{table}[h] \centering 
  \caption{Logistic regression models predicting the odds of complying across different experimental conditions. 
Column (1) presents overall compliance, while Columns (2) and (3) distinguish between compliance when the AI was incorrect and correct, respectively. Reported coefficients represent log-odds estimates, with standard errors in parentheses.} 
  \label{table: logistic compliance models} 
\begin{tabular}{@{\extracolsep{5pt}}lccc} 
\\[-1.8ex]\hline 
\hline \\[-1.8ex] 
 & \multicolumn{3}{c}{\textit{Dependent variable: Odds of Complying}} \\ 
\cline{2-4} 
\\[-1.8ex] & Overall & Helper Incorrect & Helper Correct \\ 
\\[-1.8ex] & (1) & (2) & (3)\\ 
\hline \\[-1.8ex] 
 Decision + 80\% Exp & 0.270$^{**}$ & 0.288 & 0.279$^{*}$ \\ 
  & (0.102) & (0.162) & (0.138) \\ 
  & & & \\ 
 Decision + Confidence Exp & 0.305$^{**}$ & $-$0.068 & 0.721$^{***}$ \\ 
  & (0.101) & (0.152) & (0.151) \\ 
  & & & \\ 
 Decision + self-assessing Exp & 0.268$^{**}$ & $-$0.193 & 0.803$^{***}$ \\ 
  & (0.100) & (0.151) & (0.154) \\ 
  & & & \\ 
 Age & $-$0.003 & 0.001 & $-$0.006 \\ 
  & (0.003) & (0.005) & (0.005) \\ 
  & & & \\ 
 Training Score & 0.098$^{**}$ & $-$0.077 & 0.240$^{***}$ \\ 
  & (0.032) & (0.056) & (0.041) \\ 
  & & & \\ 
 Constant (Decision-Only) & 1.412$^{***}$ & 1.585$^{***}$ & 0.912$^{*}$ \\ 
  & (0.279) & (0.468) & (0.368) \\ 
  & & & \\ 
\hline \\[-1.8ex] 
Observations & 219 & 219 & 219 \\ 
Log Likelihood & $-$508.679 & $-$362.276 & $-$407.833 \\ 
Akaike Inf. Crit. & 1,029.357 & 736.552 & 827.667 \\ 
\hline 
\hline \\[-1.8ex] 
\textit{Note:}  & \multicolumn{3}{r}{$^{*}$p$<$0.05; $^{**}$p$<$0.01; $^{***}$p$<$0.001} \\ 
\end{tabular} 
\end{table} 
\begin{table}[h] \centering 
  \caption{Logistic regression models predicting compliance when AI is incorrect. Column (1) examines compliance when the AI is overconfident, while Column (2) examines compliance when the AI correctly assesses its weakness and strength. Reported coefficients represent log-odds estimates, with standard errors in parentheses.}
  \label{tab:compliance incorrect} 
\begin{tabular}{@{\extracolsep{5pt}}lcc} 
\\[-1.8ex]\hline 
\hline \\[-1.8ex] 
 & \multicolumn{2}{c}{\textit{Dependent variable:}} \\ 
\cline{2-3} 
\\[-1.8ex] & compliance when overconfident & compliance when confidence is right \\ 
\\[-1.8ex] & (1) & (2)\\ 
\hline \\[-1.8ex] 
 ConditionDecision + 80\% Exp & 0.272 & 0.323 \\ 
  & (0.281) & (0.207) \\ 
  & & \\ 
 ConditionDecision + Confidence Exp & 1.026$^{**}$ & $-$0.498$^{*}$ \\ 
  & (0.334) & (0.195) \\ 
  & & \\ 
 ConditionDecision + self-assessing Exp & 0.632$^{*}$ & $-$0.595$^{**}$ \\ 
  & (0.300) & (0.196) \\ 
  & & \\ 
 age & 0.002 & 0.0003 \\ 
  & (0.010) & (0.006) \\ 
  & & \\ 
 trainingScore & 0.063 & $-$0.152$^{*}$ \\ 
  & (0.105) & (0.072) \\ 
  & & \\ 
 Constant & 1.134 & 1.666$^{**}$ \\ 
  & (0.881) & (0.601) \\ 
  & & \\ 
\hline \\[-1.8ex] 
Observations & 219 & 219 \\ 
Log Likelihood & $-$178.908 & $-$314.412 \\ 
Akaike Inf. Crit. & 369.817 & 640.823 \\ 
\hline 
\hline \\[-1.8ex] 
\textit{Note:}  & \multicolumn{2}{r}{$^{*}$p$<$0.05; $^{**}$p$<$0.01; $^{***}$p$<$0.001} \\ 
\end{tabular} 
\end{table}

Our self-reported measures on participants' understanding of the AI helper did not reveal any significant differences across groups (Tables \ref{table:opinion-of-ai-helper}, \ref{table:opinion-of-AI-helper2}). OpinionofAIhelper 1 to 4 indicate self-reported questions described in section \ref{sec:measures}. The willingness to pay column (Table \ref{table:opinion-of-AI-helper2}, column (2)) is another self-reported metric we asked from participants to evaluate the potential impact of different treatment conditions on reducing algorithmic aversion. Participants were asked: \textit{If you were to repeat this task but had to pay for AI suggestions from your bonus, what percentage of your bonus would you be willing to give up to receive them?}

It is possible that with a larger sample size and more detailed questions, we could gain deeper insights into how different levels of information influence users' comprehension of the AI system.
\begin{table}[!htbp] \centering 
  \caption{Linear regression models predicting participants' self-reported opinions of the AI helper. 
Columns (1), (2), and (3) correspond to different opinion measures (opinionOfAIHelper\_1, opinionOfAIHelper\_2, and opinionOfAIHelper\_3) (refer to section \ref{sec:measures}).
Reported coefficients represent the estimated effects of experimental conditions and other factors, with standard errors in parentheses.} 
  \label{table:opinion-of-ai-helper} 
\begin{tabular}{@{\extracolsep{5pt}}lccc} 
\\[-1.8ex]\hline 
\hline \\[-1.8ex] 
 & \multicolumn{3}{c}{\textit{Dependent variable:}} \\ 
\cline{2-4} 
\\[-1.8ex] & opinionOfAIHelper\_1 & opinionOfAIHelper\_2 & opinionOfAIHelper\_3 \\ 
\\[-1.8ex] & (1) & (2) & (3)\\ 
\hline \\[-1.8ex] 
 ConditionDecision + 80\% Exp & $-$13.992 & $-$32.583 & $-$47.632 \\ 
  & (51.792) & (73.817) & (61.177) \\ 
  & & & \\ 
 ConditionDecision + Confidence Exp & $-$28.992 & $-$124.577 & $-$106.734 \\ 
  & (58.224) & (82.985) & (68.775) \\ 
  & & & \\ 
 ConditionDecision + self-assessing Exp & 9.148 & $-$70.782 & $-$6.990 \\ 
  & (40.426) & (57.618) & (47.752) \\ 
  & & & \\ 
 Score & $-$9.140 & $-$59.683 & $-$45.181 \\ 
  & (43.943) & (62.631) & (51.906) \\ 
  & & & \\ 
 age & 0.379$^{**}$ & 0.213 & 0.269 \\ 
  & (0.122) & (0.174) & (0.144) \\ 
  & & & \\ 
 trainingScore & $-$1.285 & $-$2.579 & $-$0.026 \\ 
  & (1.718) & (2.449) & (2.030) \\ 
  & & & \\ 
 ConditionDecision + 80\% Exp:Score & 30.536 & 39.561 & 77.131 \\ 
  & (65.115) & (92.806) & (76.915) \\ 
  & & & \\ 
 ConditionDecision + Confidence Exp:Score & 45.187 & 164.060 & 148.271 \\ 
  & (71.556) & (101.987) & (84.523) \\ 
  & & & \\ 
 ConditionDecision + self-assessing Exp:Score & $-$3.761 & 103.908 & 30.878 \\ 
  & (50.472) & (71.935) & (59.618) \\ 
  & & & \\ 
 Constant & 68.503 & 98.811 & 80.620 \\ 
  & (36.170) & (51.552) & (42.725) \\ 
  & & & \\ 
\hline \\[-1.8ex] 
Observations & 219 & 219 & 219 \\ 
R$^{2}$ & 0.074 & 0.064 & 0.094 \\ 
Adjusted R$^{2}$ & 0.034 & 0.024 & 0.054 \\ 
Residual Std. Error (df = 209) & 20.952 & 29.862 & 24.749 \\ 
F Statistic (df = 9; 209) & 1.859 & 1.589 & 2.396$^{*}$ \\ 
\hline 
\hline \\[-1.8ex] 
\textit{Note:}  & \multicolumn{3}{r}{$^{*}$p$<$0.05; $^{**}$p$<$0.01; $^{***}$p$<$0.001} \\ 
\end{tabular} 
\end{table}

\begin{table}[!htbp] \centering 
  \caption{Linear regression models predicting participants' opinions of the AI helper and their willingness to pay for AI assistance. 
Column (1) corresponds to self-reported opinions of the AI (\textit{opinionOfAIHelper\_4}), while Column (2) examines the willingness to pay for AI-generated suggestions. 
Reported coefficients represent the estimated effects of experimental conditions and other factors, with standard errors in parentheses. } 
  \label{table:opinion-of-AI-helper2} 
\begin{tabular}{@{\extracolsep{5pt}}lcc} 
\\[-1.8ex]\hline 
\hline \\[-1.8ex] 
 & \multicolumn{2}{c}{\textit{Dependent variable:}} \\ 
\cline{2-3} 
\\[-1.8ex] & opinionOfAIHelper\_4 & Willingness to Pay \\ 
\\[-1.8ex] & (1) & (2)\\ 
\hline \\[-1.8ex] 
 ConditionDecision + 80\% Exp & $-$9.064 & 147.604$^{*}$ \\ 
  & (73.000) & (59.557) \\ 
  & & \\ 
 ConditionDecision + Confidence Exp & $-$36.460 & 23.642 \\ 
  & (85.962) & (66.954) \\ 
  & & \\ 
 ConditionDecision + self-assessing Exp & 8.750 & 62.548 \\ 
  & (56.624) & (46.487) \\ 
  & & \\ 
 Score & $-$9.981 & 110.589$^{*}$ \\ 
  & (61.965) & (50.531) \\ 
  & & \\ 
 age & 0.024 & $-$0.010 \\ 
  & (0.176) & (0.140) \\ 
  & & \\ 
 trainingScore & $-$3.315 & $-$1.958 \\ 
  & (2.421) & (1.976) \\ 
  & & \\ 
 ConditionDecision + 80\% Exp:Score & 11.584 & $-$188.551$^{*}$ \\ 
  & (91.957) & (74.877) \\ 
  & & \\ 
 ConditionDecision + Confidence Exp:Score & 50.406 & $-$28.057 \\ 
  & (106.140) & (82.285) \\ 
  & & \\ 
 ConditionDecision + self-assessing Exp:Score & $-$5.649 & $-$71.310 \\ 
  & (70.921) & (58.039) \\ 
  & & \\ 
 Constant & 82.513 & $-$55.420 \\ 
  & (50.939) & (41.593) \\ 
  & & \\ 
\hline \\[-1.8ex] 
Observations & 199 & 219 \\ 
R$^{2}$ & 0.026 & 0.067 \\ 
Adjusted R$^{2}$ & $-$0.020 & 0.027 \\ 
Residual Std. Error & 28.637 (df = 189) & 24.093 (df = 209) \\ 
F Statistic & 0.563 (df = 9; 189) & 1.662 (df = 9; 209) \\ 
\hline 
\hline \\[-1.8ex] 
\textit{Note:}  & \multicolumn{2}{r}{$^{*}$p$<$0.05; $^{**}$p$<$0.01; $^{***}$p$<$0.001} \\ 
\end{tabular} 
\end{table} 
\clearpage
\section{Survey Questions} \label{appendix:survey}
Below is the complete list of questions (in the exact order) that participants were asked to answer after the training phase and upon completing the 40 predictions.
\includepdf[pages={25-40}]{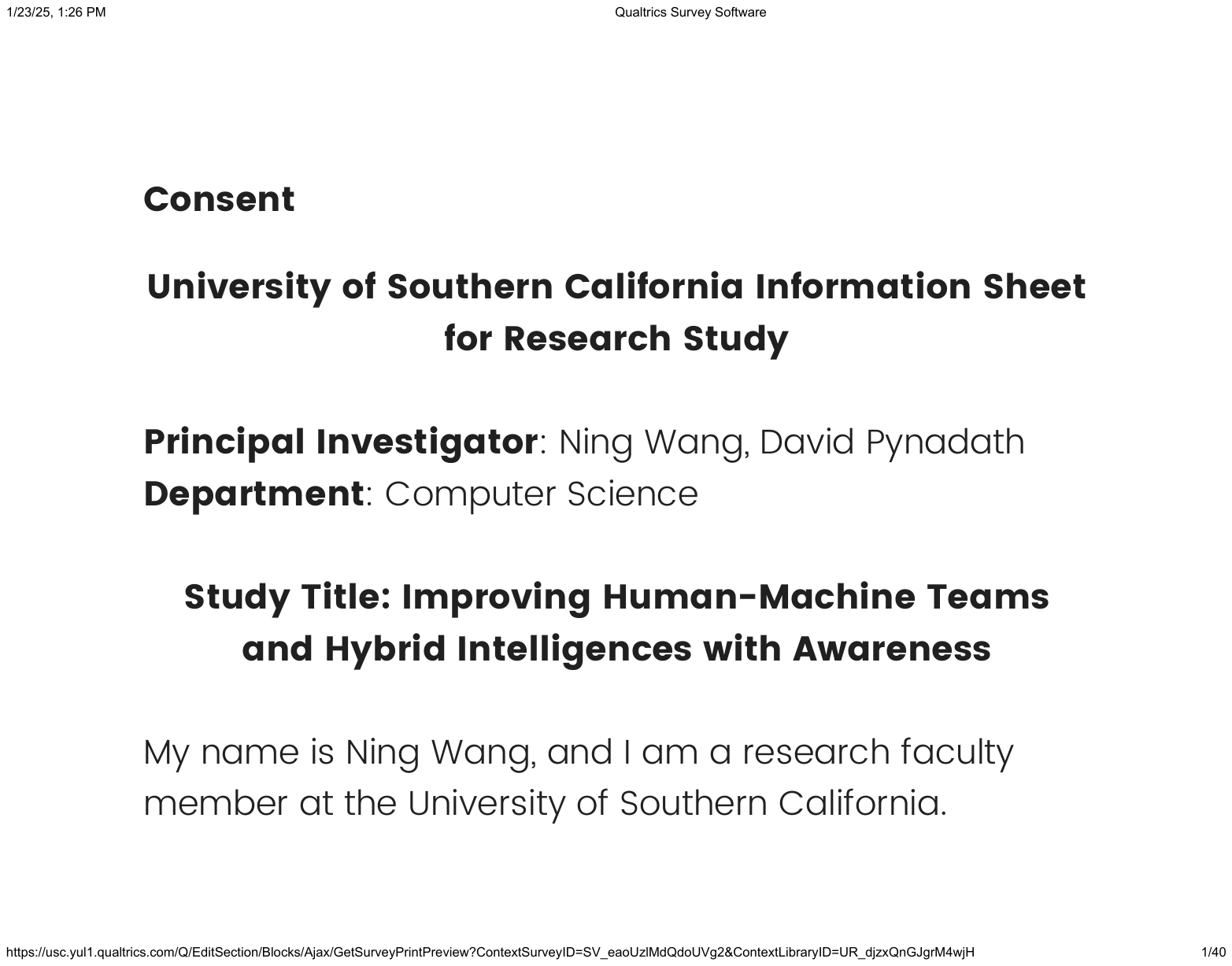}

\end{document}